\documentclass[11pt]{article}
\usepackage{makeidx}
\usepackage{amsfonts}
\usepackage{amsmath}

\newtheorem{theorem}{Theorem}
\newtheorem{acknowledgement}[theorem]{Acknowledgement}

\newtheorem{corollary}[theorem]{Corollary}

\newtheorem{lemma}[theorem]{Lemma}

\newtheorem{proposition}[theorem]{Proposition}
\newtheorem{remark}[theorem]{Remark}

\newenvironment{proof}[1][Proof]{\noindent\textbf{#1.} }{\ \rule{0.5em}{0.5em}}
\input{tcilatex}

\begin{document}

\title{ON\ THE\ GOODNESS\ OF\ \textquotedblleft QUANTUM\
BLOBS\textquotedblright\ AS SYMPLECTIC\ INVARIANT\ QUANTUM\ CELLS}
\author{Maurice de Gosson \\
Blekinge Institute of Technology\\
Department of Mathematics\\
Karlskrona 371 79, Sweden\\
e-mail \ mdegosson@web.de}
\maketitle

\begin{abstract}
We introduce a\ notion of quantum cell which is invariant under linear
canonical transformations. These cells, which we dub \textquotedblleft
quantum blobs\textquotedblright , \ have the essential and characteristic
property that their intersection with any plane of conjugate coordinates
passing through their center is an ellipse with area equal to one-half of
the quantum of action. We show that this notion is particularly suitable for
expressing various quantum mechanical properties, including a new geometric
setting for the Weyl--Wigner--Moyal formalism.
\end{abstract}

\section{Introduction}

It is customary in quantum physics and thermodynamics to \textquotedblleft
coarse grain\textquotedblright\ phase space in cubic \textquotedblleft
quantum cells\textquotedblright\ with volume $h^{n}$ ($n$ is the number of
degrees of freedom). This is consistent with the Weyl rule for the number $%
N(\lambda \leq E)$ of quantum states with energy $\lambda \leq E,$ which
says that%
\begin{equation}
N(\lambda \leq E)\overset{E\rightarrow \infty }{\sim }\frac{\func{Vol}%
(\Omega )}{h^{n}}  \label{weyl}
\end{equation}%
where $\Omega $ is the phase space volume within the corresponding energy
shell (see for instance Ozorio de Almeida \cite{Ozorio}). The integer $%
N(\lambda \leq E)$ thus appears as the number of quantum cells that can be
\textquotedblleft packed\textquotedblright\ inside $\Omega $. A major
drawback of these cubic cells is however that they have not enough
symmetries; in particular they do not enjoy any interesting invariance
properties under phase space transformations (canonical or not) and they are
therefore rather a heuristic tool than a really useful concept. In this
paper we propose a tractable substitute for these quantum cells, which has
the overwhelming advantage of being invariant under symplectic (= linear
canonical) transformations. We will call these substitutes \emph{quantum
blobs}. (We introduced a slightly different notion, under the same name, in 
\cite{physletta} to express the quantization of integrable Hamiltonian
systems in terms of properties from symplectic topology; also see \cite%
{select,paselect} for related constructions).

By definition, a quantum blob is the image of a phase space ball with radius 
$\sqrt{\hbar }$ by a (linear) symplectic transformation. (In particular,
when $n=1$, quantum blobs are just phase plane ellipses with area equal to $%
\frac{1}{2}h$). This is a \emph{simple} definition, but we will see that it
is also \emph{useful} because it will allow us to formulate various results
of phase space quantization in a both precise and concise way, and to deduce
new results. One of the main properties of quantum blobs is the following:
quantum blobs are phase space ellipsoids, but of a very special kind:

\begin{center}
\emph{The intersection of a quantum blob }$\mathbb{Q}$ \emph{by any plane
passing through its center and parallel to a plane of conjugate variables }$%
x_{j},p_{j}$ \emph{is always an ellipse with area }$\frac{1}{2}h$\emph{.}
\end{center}

\noindent Thus, quantum blobs can be interpreted as phase space sets of
\textquotedblleft minimum uncertainty\textquotedblright . This fundamental
property (in fact a more general necessary and sufficient condition) will be
proved in detail in Section \ref{two} using properties of the symplectic
group $Sp(n)$, but here is a simple dynamical proof. We may without
restricting the generality of the argument assume that the quantum blob $%
\mathbb{Q}$ is centered at the origin, so that $\mathbb{Q}$ is the image of
the phase space ball $B^{2n}(\sqrt{\hbar }):|x|^{2}+|p|^{2}\leq \hbar $ by a
linear symplectic transformation $S$. The intersection of $\mathbb{Q}$ by
any conjugate plane, say the $x_{1},p_{1}$ plane is an ellipse $\varepsilon $%
, and that ellipse is the image by $S$ of a big circle $\gamma $ of the
sphere $|x|^{2}+|p|^{2}=\hbar $. We can view $\gamma $ as a Hamiltonian
orbit determined by, say, the Hamiltonian function $H=|z|^{2}$, and the
action of that orbit (for one turn) is the area enclosed by $\gamma $, that
is 
\begin{equation*}
\oint_{\gamma }p\mathrm{d}x=\pi (\sqrt{\hbar })^{2}=\frac{1}{2}h\text{.}
\end{equation*}%
Now, the action integral along closed curves is a symplectic invariant, so
that%
\begin{equation*}
\oint_{\gamma }p\mathrm{d}x=\oint_{\varepsilon }p\mathrm{d}x=\frac{1}{2}h%
\text{;}
\end{equation*}%
since $\varepsilon $ lies in the $x_{1},p_{1}$ plane we have%
\begin{equation*}
\oint_{\varepsilon }p\mathrm{d}x=\oint_{\varepsilon }p_{1}\mathrm{d}x_{1}=%
\frac{1}{2}h
\end{equation*}%
showing hat the area enclosed by $\varepsilon $ is $\frac{1}{2}h$, as
claimed.

We urge the reader to note that this property is neither trivial, nor
obvious: if we deform a ball using a linear transformation this ball will
get stretched out and the ellipses we get by intersecting that ellipsoid by
arbitrary planes have no reason in general to have equal areas: it is not
the same thing to cut an egg in two equal parts along its long axis or its
short axis! What we actually have proven above is a linear version of
Gromov's \cite{Gromov} famous non-squeezing theorem: it is impossible to
send a ball inside a cylinder with smaller radius based on a plane of
conjugate variables.

Besides the property above quantum blobs have two other interesting
features. The first is obvious: even though their volume $h^{n}/(n!2^{n})$
is much smaller than that of a cubic cell, they can have arbitrarily large
spatial extensions (this is, by the way, the reason for which we prefer to
call our objects \textquotedblleft blobs\textquotedblright\ instead of
cells: a blob is something that can spread out); in that sense quantum blobs
are good candidates to account for EPR quantum non-locality. The other
property is that they are related to Gaussian states: let $\limfunc{Quant}%
_{0}(n)$ denote the set of all quantum blobs centered at the origin $z_{0}=0$
of phase space: $\limfunc{Quant}_{0}(n)$ is thus the orbit of the ball $%
B^{2n}(0,\sqrt{\hbar })$ under the action of the symplectic group $\limfunc{%
Sp}(n)$.The stabilizer of $B^{2n}(0,\sqrt{\hbar })$ under this action is the
unitary group $\limfunc{U}(n)$ hence we have the identification%
\begin{equation*}
\limfunc{Quant}\nolimits_{0}(n)\equiv \limfunc{Sp}(n)/\limfunc{U}(n)\text{.}
\end{equation*}%
It turn out that there exists a canonical identification between the coset
space $\limfunc{Sp}(n)/\limfunc{U}(n)$ and the space of all Wigner
transforms of Gaussians (Littlejohn \cite{Littlejohn}, \textit{p}. 265).
This suggests that there is a strong relation between quantum blobs and the
Gaussian states of quantum mechanics. That this is indeed the case, and in a
very explicit way, will be shown in Section \ref{four}: the ball $B^{2n}(0,%
\sqrt{\hbar })$ correspond to coherent states while the elements of $%
\limfunc{Quant}\nolimits_{0}(n)$ corresponds to squeezed states. The reader
should however not draw hastily the conclusion that the consideration of
quantum blobs only leads to a geometric \textquotedblleft Doppelg\"{a}%
nger\textquotedblright\ of the WWM formalism: we will see that they will
allow us to prove properties of quantum states which are not immediately
analytically obvious.

This paper is organized as follows:

\begin{itemize}
\item In Section \ref{one} we review Williamson's symplectic diagonalization
theorem and complement it with a uniqueness result modulo symplectic
rotations. we take the opportunity to study in some detail the notion of
symplectic spectrum of a symmetric positive definite matrix and deduce from
its properties a linear version of Gromov's celebrated non-squeezing theorem;

\item In Section \ref{two} we state and prove the main properties of quantum
blobs; in particular we show that the property of being a quantum blob is
\textquotedblleft hereditary\textquotedblright\ in the sense that an
ellipsoid in a symplectic subspace of\ phase space is a quantum blob if and
only if it is the intersection of a quantum blob in the total phase space
containing the given subspace;

\item In Section \ref{three} we analyze the notion of quantum blob from the
point of view of the statistical covariance matrix, so useful both in
quantum mechanics and optics. We introduce for that purpose the notion of
\textquotedblleft quantum mechanically admissible\textquotedblright\
phase-space ellipsoid;

\item In Section \ref{four} we apply the notions introduced above to the
Wigner transform of quantum states and prove that there exists a canonical
bijection between minimum uncertainty states (also called \textquotedblleft
coherent states\textquotedblright\ in quantum optics) and the set of all
quantum blobs; this justifies a posteriori their definition and
interpretation as minimum uncertainty sets. We also show that if we smooth
out a phase-space Gaussian using the coherent state associated with a
quantum blob then we obtain the Wigner transform of a mixed state.
\end{itemize}

\noindent \textbf{Notations}. The standard symplectic form on phase space $%
\mathbb{R}_{z}^{2n}=\mathbb{R}_{x}^{n}\times \mathbb{R}_{p}^{n}$ is denoted
by $\omega $:%
\begin{equation*}
\omega (z,z^{\prime })=(z^{\prime })^{T}Jz
\end{equation*}%
where $J$ is the block matrix%
\begin{equation*}
J=%
\begin{bmatrix}
0 & I \\ 
-I & 0%
\end{bmatrix}%
\text{.}
\end{equation*}%
The corresponding symplectic group is denoted by $\limfunc{Sp}(n)$:%
\begin{equation*}
S\in \limfunc{Sp}(n)\Longleftrightarrow S^{T}JS=J\Longleftrightarrow
SJS^{T}=J\text{.}
\end{equation*}%
As usual the unitary group $U(n,\mathbb{C})$ is identified with the group $%
\limfunc{U}(n)=\limfunc{Sp}(n)\cap \limfunc{O}(2n)$ of symplectic rotations
via the monomorphism%
\begin{equation*}
A+\mathrm{i}B\longmapsto 
\begin{bmatrix}
A & -B \\ 
B & A%
\end{bmatrix}%
.
\end{equation*}%
When $M$ is a positive definite symmetric matrix we will write $M>0$, and
when it is semidefinite $M\geq 0$.

The closed ball $|z-z_{0}|\leq r$ will be denoted by $B^{2n}(z_{0},r)$; when 
$z_{0}=0$ we will simply write $B^{2n}(r)$.

\section{Williamson's Theorem and Related Results\label{one}}

This Section is devoted to technical results related to Williamson's \cite%
{Will} symplectic diagonalization theorem.

\subsection{Williamson's theorem}

Let $M$ be a real $m\times m$ symmetric matrix: $M=M^{T}$. Elementary linear
algebra tells us that all the eigenvalues $\lambda _{1},\lambda
_{2},...,\lambda _{n}$ of $M$ are real, and that $M$ can be diagonalized
using an orthogonal transformation. Williamson's theorem provides us with
the symplectic variant of this result. It says that every symmetric and
positive definite matrix $M$ can be diagonalized using symplectic matrices,
and this in a very particular way:

\begin{description}
\item[\textbf{Williamson's theorem:}] \textit{For every }$2n\times 2n$%
\textit{\ matrix }$M>0$ \textit{there exists }$S\in \limfunc{Sp}(n)$\textit{%
\ such that }$M=S^{T}DS$\textit{\ with }%
\begin{equation*}
D=%
\begin{bmatrix}
\Lambda _{\omega } & 0 \\ 
0 & \Lambda _{\omega }%
\end{bmatrix}%
\text{ \ , \ }\Lambda _{\omega }=\limfunc{diag}[\lambda _{\omega
,1},...,\lambda _{\omega ,n}]\text{ \ , \ }\lambda _{\omega ,j}>0
\end{equation*}%
\textit{where the numbers }$\pm \mathrm{i}\lambda _{\omega ,j}$\textit{\ are
the eigenvalues of }$JM$\textit{.}
\end{description}

\noindent This result was initially proved by Williamson \cite{Will}, and
has been periodically rediscovered by mathematicians; Folland \cite{Folland}
or Hofer--Zehnder \cite{HZ} give \textquotedblleft modern\textquotedblright\
proofs (Hofer and Zehnder claim that the result goes back to Weierstrass).
We notice that Simon \textit{et al}. \cite{SCS} have proved a much more
general result using elementary methods.

An important observation is the following: while the diagonal entries of $%
\Lambda _{\omega }$ are uniquely determined (up to a reordering) by $M$,
there is no reason for the diagonalizing symplectic matrix $S$ to be unique.
However:

\begin{lemma}
\label{us}Assume that $S$ and $S^{\prime }$ are two elements of $\limfunc{Sp}%
(n)$ such that%
\begin{equation*}
M=(S^{\prime })^{T}DS^{\prime }=S^{T}DS.
\end{equation*}%
There exists $U\in \limfunc{U}(n)$ such that $S=US^{\prime }$.
\end{lemma}

\begin{proof}
Set $U=S(S^{\prime })^{-1}$; clearly $U$ is symplectic: $U^{T}JU=UJU^{T}=J$;
in addition $U^{T}DU=D$. We are going to show that these relations imply $%
UJ=JU$; this will prove our claim since then $U^{T}=U^{-1}$ so that $U$ is
both orthogonal and symplectic. Set $R=D^{1/2}UD^{-1/2}$; we have 
\begin{eqnarray*}
R^{T}R &=&D^{-1/2}(U^{T}DU)D^{-1/2} \\
&=&D^{-1/2}DD^{-1/2} \\
&=&I
\end{eqnarray*}%
hence $R\in $ $\limfunc{O}(2n)$. Since $J$ commutes with each power of $D$
we have (since $JU=(U^{T})^{-1}J$) 
\begin{eqnarray*}
JR &=&D^{1/2}JUD^{-1/2}=D^{1/2}(U^{T})^{-1}JD^{-1/2} \\
&=&D^{1/2}(U^{T})^{-1}D^{-1/2}J=(R^{T})^{-1}J
\end{eqnarray*}%
hence $R\in \limfunc{Sp}(n)$; $R$ is thus a symplectic rotation and
therefore $JR=RJ$. Now $U=D^{-1/2}RD^{1/2}$ and therefore%
\begin{eqnarray*}
JU &=&JD^{-1/2}RD^{1/2}=D^{-1/2}JRD^{1/2} \\
&=&D^{-1/2}RJD^{1/2}=D^{-1/2}RD^{1/2}J \\
&=&UJ
\end{eqnarray*}%
which was to be proven.
\end{proof}

When the matrix $M$ is itself symplectic, it can be diagonalized using a
symplectic \textit{rotation}:

\begin{lemma}
\label{unit}Let $S\in \limfunc{Sp}(n)$ be symmetric and positive definite.
There exists $U\in \limfunc{U}(n)$ such that%
\begin{equation}
S=U^{T}\Delta U\text{ \ \ , \ \ }\Delta =%
\begin{bmatrix}
\Lambda & 0 \\ 
0 & \Lambda ^{-1}%
\end{bmatrix}
\label{sud}
\end{equation}%
where $\Lambda =\limfunc{diag}[\lambda _{1},...,\lambda _{n}]$ and $%
0<\lambda _{1}\leq \lambda _{2}\leq \cdot \cdot \cdot \leq \lambda _{n}\leq
1 $ is a choice of $n$ first eigenvalues of $S$ (counted with their
multiplicities).
\end{lemma}

\begin{proof}
Since $S$ is symmetric and positive definite its eigenvalues occur in pairs $%
(\lambda ,1/\lambda )$ $\lambda >0$ so that if $\lambda _{1}\leq \cdot \cdot
\cdot \leq \lambda _{n}$ are $n$ eigenvalues then $1/\lambda
_{1},...,1/\lambda _{n}$ are the other $n$ eigenvalues. Let now $U$ be an
orthogonal matrix such that $S=U^{T}\Delta U$ where 
\begin{equation*}
D=\limfunc{diag}[\lambda _{1},...,\lambda _{n};1/\lambda _{1},...,1/\lambda
_{n}];
\end{equation*}%
let us prove that $U\in \limfunc{U}(n)$. Let $u_{1},...,u_{n}$ be
orthonormal eigenvectors of $U$ corresponding to the eigenvalues $\lambda
_{1},...,\lambda _{n}$. Since $SJ=JS^{-1}$ ($S$ is both symplectic and
symmetric) we have, for $1\leq k\leq n$, 
\begin{equation*}
SJu_{k}=JS^{-1}u_{k}=\frac{1}{\lambda _{j}}Ju_{k}
\end{equation*}%
hence $\pm Ju_{1},...,\pm Ju_{n}$ are the orthonormal eigenvectors of $U$
corresponding to the remaining$\ n$ eigenvalues $1/\lambda
_{1},...,1/\lambda _{n}$. Write now the $2n\times n$ matrix $%
[u_{1},...,u_{n}]$ as 
\begin{equation*}
\lbrack u_{1},...,u_{n}]=%
\begin{bmatrix}
A \\ 
B%
\end{bmatrix}%
\end{equation*}%
where $A$ and $B$ are of order $n\times n$; we have 
\begin{equation*}
\lbrack -Ju_{1},...,-Ju_{n}]=-J%
\begin{bmatrix}
A \\ 
B%
\end{bmatrix}%
=%
\begin{bmatrix}
-B \\ 
A%
\end{bmatrix}%
\end{equation*}%
hence $U$ is of the type 
\begin{equation*}
U=[u_{1},...,u_{n};-Ju_{1},...,-Ju_{n}]=%
\begin{bmatrix}
A & -B \\ 
B & A%
\end{bmatrix}%
\text{.}
\end{equation*}%
Since $U^{T}U=I$ the blocks $A$ and $B$ are such that 
\begin{equation}
AB^{T}=B^{T}A\text{ \ , \ }AA^{T}+BB^{T}=I_{n}  \label{abit}
\end{equation}%
hence $U$ is also symplectic, that is $U\in \limfunc{U}(n)$.
\end{proof}

\begin{remark}
\label{rem1}Observe that the diagonalization formula (\ref{sud}) is not a
Williamson diagonalization. However, writing%
\begin{equation}
S=(\Delta ^{1/2}U)^{T}(\Delta ^{1/2}U)  \label{sdu}
\end{equation}%
where $\Delta ^{1/2}$ is the positive square root of $\Delta $ we obtain the
Williamson diagonalization of $S$. This is because the eigenvalues of $JS$
are those of $S^{1/2}JS^{1/2}=J$ ($S^{1/2}$ is symplectic and symmetric),
and hence their moduli are all $+1$.
\end{remark}

\subsection{A property of the symplectic spectrum}

We will always use the following convention: the eigenvalues $\lambda
_{\omega ,j}$ of $JM$ will be ranked in decreasing order: 
\begin{equation}
\lambda _{\omega ,1}\geq \lambda _{\omega ,2}\geq \cdot \cdot \cdot \geq
\lambda _{\omega ,n}>0  \label{order}
\end{equation}%
and we will call the $n$-tuple%
\begin{equation*}
\limfunc{Spec}\nolimits_{\omega }(M)=(\lambda _{\omega ,1},\lambda _{\omega
,2},...,\lambda _{\omega ,n})
\end{equation*}%
the \emph{symplectic spectrum} of $M.$ The symplectic spectrum is a
symplectic invariant in the following sense: 
\begin{equation}
\limfunc{Spec}\nolimits_{\omega }(S^{T}MS)=\limfunc{Spec}\nolimits_{\omega
}(M)\text{ \ for every }S\in \limfunc{Sp}(n).  \label{specinv}
\end{equation}%
This property is of course an immediate consequence of the definition of $%
\limfunc{Spec}\nolimits_{\omega }(M)$.

A very important result is the following; it allows to compare the
symplectic spectra of two positive definite symmetric matrices.

\begin{theorem}
\label{MJ}\textit{Let }$M$ and $M^{\prime }$ be two symmetric positive
definite matrices.\textit{\ }If $M\leq M^{\prime }$ then $\limfunc{Spec}%
\nolimits_{\omega }(M)\leq \limfunc{Spec}\nolimits_{\omega }(M^{\prime })$.
\end{theorem}

\begin{proof}
We are following almost verbatim the clever argument of Giedke \textit{et al}%
. \cite{GECP} (for a proof using a variational argument see Hofer and
Zehnder \cite{HZ}). When two matrices $M$ and $M^{\prime }$ have the same
eigenvalues we will write $M\simeq M^{\prime }$. Of course $MM^{\prime
}\simeq M^{\prime }M$. We thus have to show that if the eigenvalues of $M$
are smaller or equal to those of $M^{\prime }$, then the eigenvalues of $%
(JM)^{2}$ are inferior or equal to those of $(JM^{\prime })^{2}$. The
relation $M\leq M^{\prime }$ is equivalent to $z^{T}Mz\leq z^{T}M^{\prime }z$
for every $z\in \mathbb{R}_{z}^{2n}$. Replacing $z$ par $(JM^{1/2})z$ we
thus have%
\begin{equation*}
z^{T}(JM^{1/2})^{T}M^{\prime }(JM^{1/2})z\leq z^{T}(JM^{1/2})^{T}M(JM^{1/2})z
\end{equation*}%
that is, equivalently,%
\begin{equation}
M^{1/2}JMJM^{1/2}\leq M^{1/2}JM^{\prime }JM^{1/2}  \label{jm1}
\end{equation}%
(note that both sides of (\ref{jm1}) are symmetric and positive definite).
Similarly, replacing $z$ by $(JM^{\prime 1/2})z$ in $z^{T}M^{\prime }z\leq
z^{T}Mz$ we get 
\begin{equation}
M^{\prime 1/2}JMJM^{\prime 1/2}\leq M^{\prime 1/2}JM^{\prime }JM^{\prime 1/2}%
\text{.}  \label{jm2}
\end{equation}%
Noting that we have%
\begin{eqnarray*}
M^{1/2}JM^{\prime }JM^{1/2} &\simeq &MJM^{\prime }J\text{ \ and} \\
M^{\prime 1/2}JMJM^{\prime 1/2} &\simeq &M^{\prime }JMJ\simeq MJM^{\prime }J
\end{eqnarray*}%
we can rewrite (\ref{jm1}) et (\ref{jm2}) as 
\begin{eqnarray*}
M^{1/2}JMJM^{1/2} &\leq &MJM^{\prime }J\text{ \ and } \\
MJM^{\prime }J &\leq &M^{\prime 1/2}JM^{\prime }JM^{\prime 1/2}
\end{eqnarray*}%
and hence 
\begin{equation*}
M^{1/2}JMJM^{1/2}\leq M^{\prime 1/2}JM^{\prime }JM^{\prime 1/2}.
\end{equation*}%
Since we have 
\begin{equation*}
M^{1/2}JMJM^{1/2}\simeq (MJ)^{2}\text{ \ , \ }M^{\prime 1/2}JM^{\prime
}JM^{\prime 1/2}\simeq (M^{\prime }J)^{2}
\end{equation*}%
we finally get $(MJ)^{2}\leq (M^{\prime }J)^{2}$, which was to be proven.
\end{proof}

This result has the following consequence from which one can easily derive
Gromov's non-squeezing theorem \cite{Gromov} in the linear case:

\begin{corollary}
\label{cocoon}There exists $S\in \limfunc{Sp}(n)$ sending an ellipsoid $%
\mathbb{B}:z^{T}Mz\leq 1$ into an ellipsoid $\mathbb{B}^{\prime
}:z^{T}M^{\prime }z\leq 1$ if and only if $\limfunc{Spec}\nolimits_{\omega
}(M)\geq \limfunc{Spec}\nolimits_{\omega }(M^{\prime })$.
\end{corollary}

\begin{proof}
The condition $\limfunc{Spec}\nolimits_{\omega }(M)\geq \limfunc{Spec}%
\nolimits_{\omega }(M^{\prime })$ is necessary: assume that there exists $%
S\in \limfunc{Sp}(n)$ such that $S(\mathbb{B})\subset \mathbb{B}^{\prime };$
then 
\begin{equation*}
(S^{-1}z)^{T}MS^{-1}z\geq z^{T}M^{\prime }z
\end{equation*}%
for all $z$ and hence $(S^{-1})^{T}MS^{-1}\geq M^{\prime }$. It follows from
Theorem \ref{MJ} that%
\begin{equation*}
\limfunc{Spec}\nolimits_{\omega }((S^{-1})^{T}MS^{-1})\geq \limfunc{Spec}%
\nolimits_{\omega }(M^{\prime })
\end{equation*}%
hence $\limfunc{Spec}\nolimits_{\omega }(M)\geq \limfunc{Spec}%
\nolimits_{\omega }(M^{\prime })$ taking into account the symplectic
invariance of $\limfunc{Spec}\nolimits_{\omega }(M)$. Let us prove the
sufficiency. Williamson's theorem allows us to choose $S_{1},S_{2}\in 
\limfunc{Sp}(n)$ such that%
\begin{eqnarray*}
S_{1}(\mathbb{B}) &:&\sum \lambda _{\omega ,j}(x_{j}^{2}+p_{j}^{2})\leq 1 \\
S_{2}(\mathbb{B}^{\prime }) &:&\sum \lambda _{\omega ,j}^{\prime
}(x_{j}^{2}+p_{j}^{2})\leq 1\text{.}
\end{eqnarray*}%
The condition $\limfunc{Spec}\nolimits_{\omega }(M)\geq \limfunc{Spec}%
\nolimits_{\omega }(M^{\prime })$ implies that $S_{1}(\mathbb{B})\subset
S_{2}(\mathbb{B}^{\prime })$ and hence $S(\mathbb{B})\subset \mathbb{B}%
^{\prime }$ with $S=(S_{2})^{-1}S_{1}$.
\end{proof}

\section{Properties of Quantum Blobs\label{two}}

Let us begin by proving two preliminary results which show that quantum
blobs can be obtained using only simple symplectic transformations, such as
translations, symplectic rotations and rescalings.

\subsection{Preliminaries}

A quantum blob in $\mathbb{R}_{z}^{2n}$ is the image of a ball $B^{2n}(z_{0},%
\sqrt{\hbar })$ by some $S\in \limfunc{Sp}(n)$. We can reformulate this
definition in terms of affine symplectic transformations. Let $T(z_{0})$ be
the translation operator $z\longmapsto z_{0}$. In view of the obvious
intertwining property 
\begin{equation*}
S\circ T(z_{0})=T(Sz_{0})\circ S
\end{equation*}%
valid for all $S\in \limfunc{Sp}(n)$ and all $z_{0}$ we have%
\begin{equation}
S(B^{2n}(z_{0},\sqrt{\hbar }))=T(Sz_{0})\circ S(B^{2n}(\sqrt{\hbar }))\text{.%
}  \label{inter}
\end{equation}%
A quantum blob is thus the image of the ball $B^{2n}(\sqrt{\hbar })$
centered at $0$ by an element of the affine (or inhomogeneous) symplectic
group $\limfunc{ISp}(n)$ ($\limfunc{ISp}(n)$ consists of all $S\circ T$ (or $%
T\circ S$), $S\in \limfunc{Sp}(n)$, $T$ a translation).

The following result complements this description:

\begin{proposition}
Every quantum blob can be obtained from $B^{2n}(\sqrt{\hbar })$ by a
symplectic rescaling $(x_{j},p_{j})\longmapsto (\lambda
_{j}x_{j},p_{j}/\lambda _{j})$ ($\lambda _{j}>0$, $j=1,2,...,n$) followed by
a symplectic rotation and a translation.
\end{proposition}

\begin{proof}
In view of (\ref{inter}) it is sufficient to assume that $\mathbb{Q}%
=S(B^{2n}(\sqrt{\hbar }))$; the set $\mathbb{Q}$ thus consists of all $z$
such that $z^{T}(SS^{T})^{-1}z\leq \hbar $. In view of Lemma \ref{unit}
there exists $U\in \limfunc{U}(n)$ such that $SS^{T}=U^{T}\Delta U$ where $%
\Delta $ is a diagonal matrix of the type 
\begin{equation*}
\Delta =%
\begin{bmatrix}
\Lambda & 0 \\ 
0 & \Lambda ^{-1}%
\end{bmatrix}%
.
\end{equation*}%
Writing $U^{T}\Delta U=(\Delta ^{1/2}U)^{T}(\Delta ^{1/2}U)$ (formula (\ref%
{sdu}) in Remark \ref{rem1}) the inequality $z^{T}(SS^{T})^{-1}z\leq \hbar $
is equivalent to $|\Delta ^{1/2}Uz|^{2}\leq \hbar $ and hence $\mathbb{Q}%
=U^{T}\Delta ^{1/2}(B^{2n}(\sqrt{\hbar }))$. Now $U^{T}\in \limfunc{U}(n)$
and $\Delta \in \limfunc{Sp}(n)$ is such that if $z^{\prime }=\Delta ^{1/2}z$
then $(x_{j}^{\prime },p_{j}^{\prime })=(\lambda _{j}x_{j},p_{j}/\lambda
_{j})$ for numbers $\lambda _{j}>0$, $j=1,2,...,n$.
\end{proof}

As observed in the Introduction the quantum blobs centered at the origin can
be identified with elements of the coset space $\limfunc{Sp}(n)/\limfunc{U}%
(n)$. Since in view of formula (\ref{inter}) the set $\limfunc{Quant}(n)$ of
all quantum blobs can be obtained by letting the inhomogeneous symplectic
group $\limfunc{ISp}(n)$ act on the ball $B^{2n}(\sqrt{\hbar })$ we get the
identification%
\begin{equation*}
\limfunc{Quant}(n)=\limfunc{ISp}(n)/\limfunc{U}(n)\text{.}
\end{equation*}%
The dimension of $\limfunc{ISp}(n)$ and of $\limfunc{U}(n)$ being
respectively $n(2n+1)+2n$ and $n^{2}$ it follows that 
\begin{equation*}
\dim \limfunc{Quant}(n)=n(n+3)\text{.}
\end{equation*}%
For instance $\dim \limfunc{Quant}(1)=4$: given a disk in the phase plane we
need two parameters to move it around in the phase plane using translations,
one parameter to dilate it without changing its area and one parameter
(angle!) to rotate it.

\begin{remark}
The group $\limfunc{ISp}(n)$ being arcwise connected, so is $\limfunc{Quant}%
(n)$.
\end{remark}

\subsection{The characteristic property of a quantum blob}

We will denote phase space ellipsoids by the \textquotedblleft blackboard
bold\textquotedblright\ letter $\mathbb{B}$ and quantum blobs by $\mathbb{Q}$%
.

As already emphasized in the Introduction, a quantum blob has the property
that if we cut it through its center by a plane which is parallel to a plane
of conjugate coordinates $x_{j},p_{j}$ we will always obtain an ellipse with
area $\frac{1}{2}h$. In fact, we have a more precis result, which will
produce a necessary and sufficient condition for an ellipsoid to be a
quantum blob. Before we state and prove this condition let us introduce some
terminology:

\begin{itemize}
\item A \emph{symplectic plane} is any two-dimensional linear subspace $%
\mathbb{P}$ of $\mathbb{R}_{z}^{2n}$ on which the symplectic form is
non-degenerate; in other words $\mathbb{P}$ is a symplectic vector space
when equipped with the restriction of $\omega $.

\item An \emph{affine symplectic plane} $\mathbb{P}_{z_{0}}$ is the image of
a symplectic plane $\mathbb{P}$ by the phase-space translation $z\longmapsto
z+z_{0}$. (It is, in particular, a symplectic submanifold of $\mathbb{R}%
_{z}^{2n}$.)
\end{itemize}

\begin{theorem}
\label{un}A phase space ellipsoid $\mathbb{B}$ with center $z_{0}$ is a
quantum blob if and only if its intersection with any affine symplectic
plane $\mathbb{P}_{z_{0}}$ is an ellipse with area $\frac{1}{2}h$.
\end{theorem}

\begin{proof}
It is of course no restriction to choose $z_{0}=0$. (\textit{i}) Assume that 
$\mathbb{Q}$ is a quantum blob: $\mathbb{Q}=S(B^{2n}(\sqrt{\hbar }))$ for
some $S\in \limfunc{Sp}(n)$. Let $\mathbb{P}$ be a symplectic plane; the set 
\begin{equation*}
S^{-1}(\mathbb{Q}\cap \mathbb{P})=B^{2n}(\sqrt{\hbar })\cap S^{-1}(\mathbb{P}%
)
\end{equation*}%
is the disk $B^{2}(\sqrt{\hbar })$ in $S^{-1}(\mathbb{P})$ and has thus area 
$\frac{1}{2}h$. Since the restriction $S_{|\mathbb{P}^{\prime }}$ of $S$ to $%
\mathbb{P}^{\prime }=S^{-1}(\mathbb{P})$ is symplectic, and hence area
preserving, we have 
\begin{equation*}
\limfunc{Area}(S(B^{2n}(\sqrt{\hbar }))\cap \mathbb{P})=\limfunc{Area}(S_{|%
\mathbb{P}^{\prime }}(B^{2}(\sqrt{\hbar })))=\frac{1}{2}h
\end{equation*}%
which was to be proven [one can give an alternative proof of (\textit{i})
using Remark \ref{rem1} following Lemma \ref{unit}]. (\textit{ii}) Suppose
conversely that $\mathbb{B}$ is an ellipsoid centered at the origin and such
that 
\begin{equation*}
\limfunc{Area}(\mathbb{B}\cap \mathbb{P}\mathcal{)=}\frac{1}{2}h
\end{equation*}%
for every symplectic plane $\mathbb{P}$. Since $S(\mathbb{B}\cap \mathbb{P}%
\mathcal{)=}S(\mathbb{B})\cap S(\mathbb{P})$ and since $\limfunc{Sp}(n)$
acts transitively on symplectic planes we have 
\begin{equation*}
\limfunc{Area}(S(\mathbb{B})\cap \mathbb{P}\mathcal{)}=\frac{1}{2}h
\end{equation*}%
for every $S\in \limfunc{Sp}(n)$. Let $M>0$be such that the ellipsoid $%
\mathbb{B}$ is defined by the inequality $z^{T}Mz\leq 1$; in view of
Williamson's theorem we can find $S\in \limfunc{Sp}(n)$ such that 
\begin{equation*}
M=S^{T}DS\text{ \ , \ }D=%
\begin{bmatrix}
\Lambda _{\omega } & 0 \\ 
0 & \Lambda _{\omega }%
\end{bmatrix}%
\end{equation*}%
where $\Lambda _{\omega }=\limfunc{diag}g[\lambda _{\omega ,1},...,\lambda
_{\omega ,n}]$. The ellipsoid $S^{-1}(\mathbb{B})$ is thus defined by $%
z^{T}Dz\leq 1$, that is%
\begin{equation*}
\sum_{j=1}^{n}\lambda _{\omega ,j}(x_{j}^{2}+p_{j}^{2})\leq 1\text{.}
\end{equation*}%
It follows that the intersections of $S^{-1}(\mathbb{B})$ by the symplectic
coordinate planes $\mathbb{P}_{j}=\{z:z_{k}=0;k\neq j\}$ are the disks 
\begin{equation*}
\mathbb{D}_{j}:\lambda _{\omega ,j}(x_{j}^{2}+p_{j}^{2})\leq 1
\end{equation*}%
hence the equality%
\begin{equation*}
\limfunc{Area}\mathbb{D}_{j}=\frac{\pi }{\lambda _{\omega ,j}}=\frac{1}{2}h
\end{equation*}%
implies that we must have $\lambda _{\omega ,j}=1/\hbar $ for $j=1,2,...,n$
and $S^{-1}(\mathbb{B})$ is thus the ball $B^{2n}(\sqrt{\hbar })$ so that $%
\mathbb{B}=S(B^{2n}(\sqrt{\hbar })$.
\end{proof}

Theorem \ref{un} has the following straightforward consequence which is a
linear version of Gromov's famous non-squeezing theorem \cite{Gromov}:

\begin{corollary}
\label{coco}The orthogonal projection of a quantum blob $\mathbb{Q}$ on any
symplectic affine plane $\mathbb{P}_{z_{1}}$ has an area at lest equal to $%
\frac{1}{2}h$.
\end{corollary}

\begin{proof}
Let $\mathbb{P}_{z_{0}}$ be the affine plane parallel to $\mathbb{P}_{z_{1}}$
and passing through the center of $\mathbb{Q}$. That plane is also an affine
symplectic plane hence $\mathbb{Q}\cap \mathbb{P}_{z_{0}}$ has area $\frac{1%
}{2}h$. The orthogonal projection of $\mathbb{Q}$ on $\mathbb{P}_{z_{1}}$
contains that of $\mathbb{Q}\cap \mathbb{P}_{z_{0}}$ and is hence superior
or equal to $\frac{1}{2}h$.
\end{proof}

\subsection{Extension to symplectic subspaces}

The observant reader will have noticed that the main ingredient in the proof
of the necessary condition in Theorem \ref{un} was the following property:

\begin{center}
\emph{The restriction of a symplectic transformation to a symplectic plane }$%
\mathbb{P}$\emph{\ is area-preserving.}
\end{center}

This suggests that we might go one step further and study the intersection
of quantum blobs by arbitrary symplectic affine subspaces passing through
their center. This will allow us to associate to any quantum blob in $%
\mathbb{R}_{z}^{2n}$ a sort of \textquotedblleft
hierarchy\textquotedblright\ of quantum blobs with smaller dimensions by
cutting the blob with symplectic (affine) spaces with smaller and smaller
dimensions.

Before we proceed to do so, recall that:

\begin{itemize}
\item A \emph{symplectic subspace} is a linear subspace $\mathbb{V}$ of $%
\mathbb{R}_{z}^{2n}$ which is a symplectic vector space when equipped with
the restriction of $\omega $. An \emph{affine symplectic subspace} is
obtained by translation in $\mathbb{R}_{z}^{2n}$ of a symplectic subspace.

\item A \emph{symplectic basis}\textit{\ }of a symplectic space $\mathbb{V}$
is a basis $(e_{1},...,e_{m};f_{1},...,f_{m})$ such that $\omega
(f_{j},e_{k})=\delta _{jk}$ and $\omega (e_{j},e_{k})=\omega (f_{j},f_{k})=0$
for $1\leq j,k\leq m$.

\item An \emph{orthosymplectic basis }of a symplectic subspace is a basis
which is both symplectic and orthogonal with respect to the scalar product
determined by $\omega $.
\end{itemize}

A symplectic subspace always has even dimension $2m$ due to the
non-degeneracy of the symplectic form. Given two symplectic subspaces $%
\mathbb{V}$ and $\mathbb{V}^{\prime }$ of same dimension there exists a
linear mapping $S_{\mathbb{V}^{\prime },\mathbb{V}}:\mathbb{V}%
\longrightarrow \mathbb{V}^{\prime }$ such that%
\begin{equation*}
\omega _{\mathbb{V}^{\prime }}(S_{\mathbb{V}^{\prime },\mathbb{V}}v,S_{%
\mathbb{V}^{\prime },\mathbb{V}}v^{\prime })=\omega _{\mathbb{V}%
}(v,v^{\prime })
\end{equation*}%
for every $v\in \mathbb{V}$; $\omega _{\mathbb{V}^{\prime }}$ and $\omega _{%
\mathbb{V}}$ are the restrictions of the symplectic form $\omega $ to $%
\mathbb{V}$ and $\mathbb{V}^{\prime }$, respectively (to construct such a $%
S_{\mathbb{V}^{\prime },\mathbb{V}}$ it suffices to choose symplectic bases
in $\mathbb{V}$ and $\mathbb{V}^{\prime }$). We will say that $S_{\mathbb{V}%
^{\prime },\mathbb{V}}$ is as symplectic mapping. Note that it is volume
preserving when $\mathbb{V}$ and $\mathbb{V}^{\prime }$ are equipped with
their canonical volume elements. When $\mathbb{V}=\mathbb{V}^{\prime }$ the
symplectic mappings $\mathbb{V}\longrightarrow \mathbb{V}$ form a group, the
symplectic group $\limfunc{Sp}(\mathbb{V})$. Using orthosymplectic bases of $%
\mathbb{V}$ and $\mathbb{V}^{\prime }$ one can as well construct mappings $%
U_{\mathbb{V}^{\prime },\mathbb{V}}:\mathbb{V}\longrightarrow \mathbb{V}%
^{\prime }$ that are both symplectic and orthogonal (orthogonality referring
here to the fact that $U_{\mathbb{V}^{\prime },\mathbb{V}}$ takes one
symplectic basis of $\mathbb{V}$ to an orthogonal basis of $\mathbb{V}%
^{\prime }$. The following Lemma summarizes two results we shall need:

\begin{lemma}
\label{elva}(i) The unitary group $\limfunc{U}(n)$ (and hence $\limfunc{Sp}%
(n)$) acts transitively on the manifold of all symplectic subspaces of $%
\mathbb{R}_{z}^{2n}$ with same dimension; (ii) Every symplectic mapping $S_{%
\mathbb{V}^{\prime },\mathbb{V}}:\mathbb{V}\longrightarrow \mathbb{V}%
^{\prime }$ is the restriction to $\mathbb{V}$ of an element of $\limfunc{Sp}%
(n)$. In particular, for every $S_{\mathbb{V}}\in \limfunc{Sp}(\mathbb{V})$
there exists $S\in \limfunc{Sp}(n)$ such that $S_{\mathbb{V}}=S_{|\mathbb{V}%
} $.
\end{lemma}

\begin{proof}
(\textit{i}) Let $\mathbb{V}$ and $\mathbb{V}^{\prime }$ be two symplectic
subspaces of $\mathbb{R}_{z}^{2n}$ such that $\dim \mathbb{V}=\dim \mathbb{V}%
^{\prime }=2k$ and let 
\begin{equation*}
\mathcal{S}_{2k}=(e_{1},...,e_{m};f_{1},...,f_{m})\text{ \ , \ }\mathcal{S}%
_{2k}^{\prime }=(e_{1}^{\prime },...,e_{m}^{\prime };f_{1}^{\prime
},...,f_{m}^{\prime })
\end{equation*}%
be orthosymplectic bases of $\mathbb{V}$ and $\mathbb{V}^{\prime }$,
respectively. Completing these bases into full symplectic bases $\mathcal{S}%
_{2n}$ and $\mathcal{S}_{2n}^{\prime }$ of $\mathbb{R}_{z}^{2n}$, the linear
mapping $S:\mathbb{R}_{z}^{2n}\longrightarrow \mathbb{R}_{z}^{2n}$defined by 
$S(\mathcal{S}_{2n})=\mathcal{S}_{2n}^{\prime }$ is symplectic and such that 
$S(\mathbb{V})=\mathbb{V}^{\prime }$; the restriction of $S$ to $\mathbb{V}$
is orthogonal since $S(\mathcal{S}_{2m})=\mathcal{S}_{2m}^{\prime }$ and the
bases $\mathcal{S}_{2m}$, $\mathcal{S}_{2m}^{\prime }$ are unitary. (\textit{%
ii}) Let $S_{\mathbb{V}^{\prime },\mathbb{V}^{\prime }}$ be a symplectic
mapping of $\mathbb{V}\longrightarrow \mathbb{V}^{\prime }$. Choose a
symplectic basis $\mathcal{S}_{2k}$ of $\mathbb{V}$ as above; then $\mathcal{%
S}_{2m}^{\prime }=(e_{1}^{\prime },...,e_{m}^{\prime };f_{1}^{\prime
},...,f_{m}^{\prime })$ with $e_{j}^{\prime }=S_{\mathbb{V}^{\prime },%
\mathbb{V}^{\prime }}(e_{j})$, $f_{j}^{\prime }=S_{\mathbb{V}^{\prime },%
\mathbb{V}}(f_{j})$, $1\leq j\leq m$, is also a symplectic basis of $\mathbb{%
V}$. Let us now again complete $\mathcal{S}_{2m}$ and $\mathcal{S}%
_{2m}^{\prime }$ into full symplectic bases $\mathcal{S}_{2n}$ and $\mathcal{%
S}_{2n}^{\prime }$ of $\mathbb{R}_{z}^{2n}$ and define $S\in \limfunc{Sp}(n)$
by $S(e_{j})=e_{j}^{\prime }$, $S(f_{j})=f_{j}^{\prime }$, $1\leq j\leq n$.
The restriction of $S$ to $\mathbb{V}$ is just $S_{\mathbb{V}^{\prime },%
\mathbb{V}^{\prime }}$.
\end{proof}

\begin{proposition}
(i) The intersection of a quantum blob $\mathbb{Q}=S(B^{2n}(z_{0},\sqrt{%
\hbar }))$ with any affine symplectic subspace $\mathbb{V}_{z_{0}}=T(z_{0})%
\mathbb{V}$ of $\mathbb{R}_{z}^{2n}$ is a quantum blob in $\mathbb{V}%
_{z_{0}} $, that is the image of a ball of $\mathbb{V}_{z_{0}}$ with radius $%
\sqrt{\hbar }$ by an element of the symplectic group of the symplectic space 
$\mathbb{V}$. (ii) Conversely, any quantum blob $\mathbb{Q}_{0}$ in $\mathbb{%
V}_{z_{0}}$ is the intersection of a quantum blob in $\mathbb{R}_{z}^{2n}$
with $\mathbb{V}_{z_{0}}$.
\end{proposition}

\begin{proof}
(\textit{i}) It is a straightforward adaptation of the proof of the
necessary condition in Theorem \ref{un}. It again suffices to assume $%
z_{0}=0 $. Let $\mathbb{V}$ be a symplectic subspace, $\dim \mathbb{V}=2m$;
then 
\begin{equation*}
S^{-1}(\mathbb{Q}\cap \mathbb{V})=B^{2n}(\sqrt{\hbar })\cap S^{-1}(\mathbb{V}%
)
\end{equation*}%
is the ball $B^{2m}(\sqrt{\hbar })$ in $S^{-1}(\mathbb{V})$. The restriction 
$S_{|\mathbb{V}^{\prime }}$ of $S$ to $\mathbb{V}^{\prime }=S^{-1}(\mathbb{V}%
)$ being symplectic, the set 
\begin{equation*}
S_{|\mathbb{V}^{\prime }}(B^{2m}(\sqrt{\hbar }))=S(B^{2n}(\sqrt{\hbar }%
))\cap \mathbb{V}
\end{equation*}%
is a quantum blob in $\mathbb{V}$. (\textit{ii}) Let $\mathbb{Q}_{0}=S_{%
\mathbb{V}}(B^{2m}(\sqrt{\hbar }))$ and choose $S\in \limfunc{Sp}(n)$ such
that $S_{|\mathbb{V}}=S_{\mathbb{V}}$ (Lemma \ref{elva}, (\textit{ii})). Let 
$\mathbb{Q}=S(B^{2n}(\sqrt{\hbar }))$; we have 
\begin{equation*}
\mathbb{Q}\cap S(\mathbb{V})=S(B^{2n}(\sqrt{\hbar })\cap \mathbb{V})=\mathbb{%
Q}_{0}\text{.}
\end{equation*}
\end{proof}

\begin{remark}
One might wonder whether it could be possible to find a ball $B^{2n}(r)$
with radius $r$ smaller than $\sqrt{\hbar }$ and $S\in \limfunc{Sp}(n)$ such
that the intersection of $S(B^{2n}(r))$ with a symplectic subspace plane is
a quantum blob. Such a possibility is however ruled out by Theorem \ref{un}.
\end{remark}

This discussion can be extended \textit{mutatis mutandis} to the general
case: every quantum blob of an affine symplectic subspace is the
intersection of a quantum blob in the whole space with that subspace. This
result can be interpreted in the following way: assume that a Hamiltonian
system consists of two parts, Alice and Bob; each of these parts has its own
phase space, which is equipped with the restriction of the overall
symplectic form to each part. The discussion above says that a quantum blob
in the whole phase space is perceived as an own private quantum blob by
Alice as well as by Bob. It would certainly be interesting to pursue this
analysis from the point of view of the so important entangled states of
quantum mechanics; we hope to come back to this issue in a not too distant
future.

Let us end this section with the following observation. Planck's constant
does of course not play any particular role in all what we have done and can
therefore be replaced by any number $r>0$. The images of arbitrary balls by
symplectic transformations could perhaps be used as measures of the
information contained in parts of Hamiltonian systems along the lines
suggested in the thesis of Wolf \cite{Wolf} where a similar situation is
discussed from the point of view of the BBGKY hierarchy.

\section{Admissible Ellipsoids and Quantum Uncertainty\label{three}}

What about ellipsoids containing quantum blobs? It turns out that these also
are of genuine interest (we will use them copiously in our study of
uncertainty and of the WWM formalism).

We will call a phase space ellipsoid containing a quantum blob a \emph{%
quantum mechanically} \emph{admissible ellipsoid} or, for short: \emph{%
admissible ellipsoid}.

\subsection{Properties of admissible ellipsoids}

The following property of admissible ellipsoids is very simple and
\textquotedblleft obvious\textquotedblright ; the proof of its first part
however requires Williamson's theorem, and the proof of the second part the
highly non-trivial Theorem \ref{MJ}!

\begin{proposition}
(i) A phase space ellipsoid $\mathbb{B}$ is quantum mechanically admissible
if and only if its intersection with every affine symplectic plane $\mathbb{P%
}$ passing through its center has area at least $\frac{1}{2}h$. (ii) An
admissible phase space ellipsoid $\mathbb{B}:z^{T}Mz\leq 1$ can be sent in
another phase space ellipsoid $\mathbb{B}^{\prime }:z^{T}M^{\prime }z\leq 1$
by an (affine) symplectic transformation if and only if $\limfunc{Spec}%
\nolimits_{\omega }(M^{\prime })\leq \limfunc{Spec}\nolimits_{\omega }(M)$.
\end{proposition}

\begin{proof}
(\textit{i}) That the condition is necessary is indeed obvious: if $\mathbb{B%
}$ contains a quantum blob $\mathbb{Q}$ then every plane through its center
will also cut $\mathbb{Q}$, and if this plane is affine symplectic then the
area of this section will be $\frac{1}{2}h$. The hard part is to prove that
it is also sufficient. Here is why: while it is true that if every section $%
\mathbb{B}\cap \mathbb{P}$ has area $\geq \frac{1}{2}h$ then it will contain
a smaller ellipse with area $\frac{1}{2}h,$ it is not clear that these
ellipses are the \textquotedblleft cuts\textquotedblright\ by symplectic
planes of a same quantum blob contained in $\mathbb{B}$. That it is actually
true is a consequence of Williamson's theorem: choose $S\in \limfunc{Sp}(n)$
such that 
\begin{equation*}
S^{-1}(\mathbb{B}):\sum_{j=1}^{n}\frac{1}{R_{j}^{2}}(x_{j}^{2}+p_{j}^{2})%
\leq 1
\end{equation*}%
(the $R_{j}$ are just the numbers $1/\sqrt{\lambda _{\omega ,j}}$). If we
now cut $S^{-1}(\mathbb{B})$ with the plane $\mathbb{P}_{j}$ of coordinates $%
x_{j},p_{j}$ we get the disk 
\begin{equation*}
B^{2}(R_{j}):x_{j}^{2}+p_{j}^{2}\leq R_{j}^{2}
\end{equation*}%
The intersection $\mathbb{B}\cap S(\mathbb{P}_{j})$ is an ellipse with same
area $\pi R_{j}^{2}$ as that disk hence $\pi R_{j}^{2}\geq \frac{1}{2}h$
that is $R_{j}\geq \sqrt{\hbar }$. This implies that $S^{-1}(\mathbb{B})$
contains the ball $B^{2n}(\sqrt{\hbar })$ and hence $\mathbb{B}\supset
S(B^{2n}(\sqrt{\hbar }))$ which was to be proven. (\textit{ii}) Since the
symplectic spectrum is insensitive to phase space translations we may assume
that $\mathbb{B}$ and $\mathbb{B}^{\prime }$ have same center $z_{0}=0$. The
result then immediately follows from the Corollary \ref{cocoon} of Theorem %
\ref{MJ}.
\end{proof}

In \cite{select,paselect} we showed that the notion of symplectic capacity
can be fruitfully used to describe in a concise way quantization of
integrable systems. The notion also leads to an economical way of describing
admissible ellipsoids. Let us recall how the (Gromov) symplectic capacity $%
\underline{c}$ is defined: if $\Omega $ is a subspace of $\mathbb{R}%
_{z}^{2n} $ we denote by $R$ the supremum of all $r\geq 0$ such that there
exists a canonical transformation $f$ of $\mathbb{R}_{z}^{2n}$ sending a
phase space ball with radius $r$ inside $\Omega $. We call that $R$ (which
can be $+\infty $) the \emph{symplectic radius} of $\Omega $ and $\pi R^{2}$
its \emph{Gromov width}. The Gromov width of $\Omega $, which we will denote
by $\underline{c}(\Omega )$, is a particular and useful case of the more
general notion of symplectic capacity. A symplectic capacity $c$ on $\mathbb{%
R}_{z}^{2n}$ assigns to every subset $\Omega \subset \mathbb{R}_{z}^{2n}$ a
non-negative number or $+\infty $ and satisfies the following
properties:\bigskip

\begin{tabular}{l}
\emph{(1) If }$f:\Omega \longrightarrow \Omega ^{\prime }$\emph{\ is
canonical then }$c(f(\Omega ))\leq c(\Omega ^{\prime });$ \\ 
\\ 
\emph{(2) }$c(k\Omega )=k^{2}c(\Omega )$\emph{\ for every real number }$k$%
\emph{;} \\ 
\\ 
\emph{(3) }$c(B^{2n}(r))=\pi r^{2}=c(Z_{j}(r))$\emph{\ where }$%
Z_{j}(r):x_{j}^{2}+p_{j}^{2}\leq r^{2}$\emph{.}%
\end{tabular}%
\bigskip

Notice that (1) implies, in particular, that $c(\Omega )\leq c(\Omega
^{\prime })$\emph{\ }if $\Omega \subset \Omega ^{\prime }$\ and that $%
c(f(\Omega ))=c(\Omega )$\ for every canonical transformation $f$.

All these properties are trivial \emph{except the last}: the equality $%
c(Z_{j}(r))=\pi r^{2}$ is namely equivalent to Gromov's theorem \cite{Gromov}
which asserts that it is impossible to \textquotedblleft
squeeze\textquotedblright , using canonical transformations, a ball inside a
cylinder $Z_{j}(r)$ with smaller radius; all known proofs of Gromov's
theorem are notoriously difficult.

\begin{remark}
It is easy to check (see for instance \cite{HZ})) that the Gromov width is
the smallest of all symplectic capacities: $\underline{c}\leq c$ for every
other symplectic capacity $c$.
\end{remark}

Let $\mathbb{B}:z^{T}Mz\leq 1$ be a phase space ellipsoid; in view of
Williamson's theorem there exists $S\in \limfunc{Sp}(n)$ such that%
\begin{equation*}
S^{-1}(\mathbb{B}):\sum_{j=1}^{n}\frac{1}{R_{j}^{2}}(x_{j}^{2}+p_{j}^{2})%
\leq 1\text{;}
\end{equation*}%
the numbers $R_{j}^{2}$ are the inverses of the elements $\lambda _{\omega
,j}$ of the symplectic spectrum of $M$, so that according to the convention (%
\ref{order}) we have 
\begin{equation}
0<R_{1}\leq R_{2}\leq \cdot \cdot \cdot \leq R_{n}\text{.}  \label{rang}
\end{equation}%
An essential property (see for instance \cite{HZ}) is that%
\begin{equation}
\underline{c}(\mathbb{B})=\underline{c}_{\text{\textrm{aff}}}(\mathbb{B}%
)=\pi R_{1}^{2}  \label{cbm}
\end{equation}%
where $\underline{c}_{\text{\textrm{aff}}}$ is the affine symplectic
capacity: $\underline{c}_{\text{\textrm{aff}}}(\mathbb{B})$ is the number $%
\pi R^{2}$ where $R$ is the supremum of the radii all balls $B^{2n}(r)$ that
can be sent in $\mathbb{B}$ using translations and elements of $\limfunc{Sp}%
(n)$.

The following result characterizes admissible ellipsoids in terms of the
symplectic capacity $\underline{c}$:

\begin{proposition}
A phase space ellipsoid $\mathbb{B}$ is quantum mechanically admissible if
and only if $\underline{c}(\mathbb{B})\geq \frac{1}{2}h$.
\end{proposition}

\begin{proof}
In view of (\ref{cbm}) the condition $\underline{c}(\mathbb{B})\geq \frac{1}{%
2}h$ is equivalent to $R_{1}\geq \sqrt{\hbar }$, that is to 
\begin{equation*}
\sqrt{\hbar }\leq R_{1}\leq R_{2}\leq \cdot \cdot \cdot \leq R_{n}\text{.}
\end{equation*}%
This is equivalent to saying that $S^{-1}(\mathbb{B})$ contains the ball $%
B^{2n}(\sqrt{\hbar })$ that is $S(B^{2n}(\sqrt{\hbar }))\subset \mathbb{B}$.
\end{proof}

If $\mathbb{B}$ is a quantum blob then we have equality in the Proposition
above: $\underline{c}(\mathbb{B})=\frac{1}{2}h$ because%
\begin{equation*}
\underline{c}(\mathbb{B})=\underline{c}(S(B^{2n}(z_{0},\sqrt{\hbar }))=%
\underline{c}(B^{2n}(z_{0},\sqrt{\hbar }))\text{.}
\end{equation*}%
The converse of this property is however not true: property (3) of
symplectic capacities says that any ellipsoid $\mathbb{B}$ such that%
\begin{equation*}
B^{2n}(\sqrt{\hbar })\subset \mathbb{B}\subset Z_{j}(\sqrt{\hbar })
\end{equation*}%
($Z_{j}(\sqrt{\hbar })$ the cylinder $x_{j}^{2}+p_{j}2$ $\leq \hbar $) will
have capacity $\frac{1}{2}h$: symplectic capacities are in a sense to
imprecise to detect\ the whole symplectic spectrum of an ellipsoid. Notice,
however, that if $n=1$ then the condition $\underline{c}(\mathbb{B})=\frac{1%
}{2}h$ means that the area of $\mathbb{B}$ is $\frac{1}{2}h$ and hence a
phase plane ellipse with symplectic capacity $\frac{1}{2}h$ is just a
quantum blob: it is only in more than one degree of freedom that the notion
becomes interesting.

Here is another important observation: If an admissible ellipsoid $\mathbb{B}
$ has symplectic capacity $\frac{1}{2}h$ then it can of course contain
several distinct quantum blobs. However, and this is an important property
we will use in Section \ref{four} when dealing with the Wigner transform, is
that one can always associate a canonical concentric \textquotedblleft
quantum blob companion\textquotedblright\ to such an ellipsoid $\mathbb{B}$.
Since the symplectic capacity is invariant under translations it suffices to
check this property when $\mathbb{B}$ is centered at $z_{0}=0$. In view of
Williamson's theorem and the fact that $\sqrt{\hbar }=R_{1}$ there exists $%
S\in \limfunc{Sp}(n)$ such that 
\begin{equation}
S(\mathbb{B}):\frac{1}{\hbar }(x_{1}^{2}+p_{1}^{2})+\sum_{j=2}^{n}\frac{1}{%
R_{j}^{2}}(x_{j}^{2}+p_{j}^{2})\leq 1\text{;}  \label{sb}
\end{equation}%
since 
\begin{equation*}
\sqrt{\hbar }=R_{1}\leq R_{2}\leq \cdot \cdot \cdot \leq R_{n}
\end{equation*}%
the ball $B^{2n}(\sqrt{\hbar })$ is contained in $S(\mathbb{B})$ and hence $%
S^{-1}(B^{2n}(\sqrt{\hbar }))\subset \mathbb{B}$. The quantum blob $\mathbb{Q%
}=S^{-1}(B^{2n}(\sqrt{\hbar }))$ is unambiguously defined, because if $%
S^{\prime }$ is another element of $\limfunc{Sp}(n)$ putting $\mathbb{B}$ in
the form (\ref{sb}) then $S=US^{\prime }$ with $U\in \limfunc{U}(n)$ in view
of Lemma \ref{us} and hence%
\begin{equation*}
S^{-1}(B^{2n}(\sqrt{\hbar }))=(S^{\prime })^{-1}(B^{2n}(\sqrt{\hbar }))\text{%
.}
\end{equation*}

\subsection{Relation with the uncertainty principle\label{reluc}}

The usual quantum cells appearing in the literature are related to the
Heisenberg uncertainty principle. So are our quantum blobs, but in a more
precise way.

A generic ellipsoid $\mathbb{B}$ in $\mathbb{R}_{z}^{2n}$ can always be
viewed as the set of all $z$ such that $z^{T}Mz\leq 1$ for some $M>0$. From
a quantum mechanical point of view it will be advantageous to replace $M$ by 
$F=\frac{1}{\hbar }M$ so that $\mathbb{B}:z^{T}Fz\leq \hbar $.

Let $F>0,$ be a phase space ellipsoid. We associate to $\mathbb{B}$ the
matrix%
\begin{equation*}
\Sigma =\frac{\hbar }{2}F^{-1}
\end{equation*}%
which will be viewed as a statistical covariance (or \textquotedblleft
noise\textquotedblright ) matrix. This interpretation is motivated by the
properties of the Wigner transform that will be investigated in next Section.

\begin{proposition}
\label{bon}The ellipsoid $\mathbb{B}:z^{T}Fz\leq \hbar $ is quantum
mechanically admissible if and only if the three following equivalent
conditions are satisfied:%
\begin{equation}
\begin{tabular}{l}
(A) $\ \ \Sigma +\frac{\mathrm{i}\hbar }{2}J\text{ \ \emph{is Hermitian
positive semi-definite};}$ \\ 
(B) $\ \ F^{-1}+\mathrm{i}J\text{ \ \emph{is Hermitian positive
semi-definite.}}$ \\ 
(C) $\ \ \text{\emph{The eigenvalues of }}J\Sigma \text{ \emph{are }}\geq 
\frac{1}{2}\hbar $ \\ 
(D) \ \ $\text{\emph{The eigenvalues of }}JF\text{ \emph{are }}\leq 1$.%
\end{tabular}
\label{unce}
\end{equation}
\end{proposition}

\begin{proof}
The conditions (A), (B) and (C), (D) are trivially equivalent in view of the
definition of $F$. It thus suffices to prove that (B)$\Longleftrightarrow $%
(D). Using Williamson's theorem we can write 
\begin{equation*}
F=S^{T}DS\text{ , }S\in \limfunc{Sp}(n)\text{ , }D=%
\begin{bmatrix}
\Lambda & 0 \\ 
0 & \Lambda%
\end{bmatrix}%
\end{equation*}%
where $\Lambda =\limfunc{diag}[\lambda _{\omega ,1},...,\lambda _{\omega
,n}] $. Since $S^{T}JS=J$ condition (A) is in turn equivalent to%
\begin{equation}
D^{-1}+\mathrm{i}J\text{ \ \emph{is Hermitian positive semi-definite}\textrm{%
.}}  \label{coda}
\end{equation}%
The characteristic polynomial $P(\lambda )$ of $D^{-1}+\mathrm{i}J$ is the
product 
\begin{equation*}
P(\xi )=P_{1}(\xi )P_{2}(\xi )\cdot \cdot \cdot P_{n}(\xi )
\end{equation*}%
where $P_{j}$ is, for $j=1,2,...,n$, the polynomial 
\begin{equation*}
P_{j}(\xi )=\xi ^{2}-2\lambda _{\omega ,j}^{-1}\xi +\lambda _{\omega
,j}^{-2}-1\text{.}
\end{equation*}%
The roots of $P$ are the numbers $\xi _{j}=\pm 1+(1/\lambda _{\omega ,j})$
and we have $\xi _{j}\geq 0$ if and only if $\lambda _{\omega ,j}\leq 1$.
\end{proof}

The Proposition above says that:

\begin{center}
\emph{An ellipsoid is admissible if and only if its associated covariance
matrix satisfies the Heisenberg uncertainty principle (\ref{unce}).}
\end{center}

Conditions (\ref{unce}) are in fact symplectically invariant statements of
Heisenberg's uncertainty principle; see Arvind \textit{et al.} see \cite%
{ADMS} and Simon \textit{et al.} \cite{SSM} (Trifonov \cite{Trifonov} gives
an interesting discussion, containing a comprehensive list of references, of
various generalizations of the uncertainty principle). For instance when $%
n=1 $ the covariance matrix is%
\begin{equation}
\Sigma =%
\begin{bmatrix}
\sigma _{x}^{2}\smallskip & \rho \sigma _{x}\sigma _{p} \\ 
\rho \sigma _{x}\sigma _{p} & \sigma _{p}^{2}%
\end{bmatrix}%
\text{ \ \ with \ }-1\leq \rho \leq 1  \label{code}
\end{equation}%
and the eigenvalues of $J\Sigma $ are 
\begin{equation*}
\mu =\pm \mathrm{i}\sigma _{x}\sigma _{p}\sqrt{1-\rho ^{2}}
\end{equation*}%
hence $B$ is quantum-mechanically admissible if and only if we have%
\begin{equation}
\sigma _{x}\sigma _{p}\sqrt{1-\rho ^{2}}\geq \frac{\hbar }{2}.  \label{haha}
\end{equation}

Let us end this Section by briefly discussing the notion of \textit{squeezed
state}. Squeezed states are of a particular interest in quantum optics
because they allow to beat the quantum limit in optical measurements (Walls
and Milburn \cite{WM}). Assume that $n=1$; a quantum state with covariance
matrix 
\begin{equation*}
\Sigma =%
\begin{bmatrix}
\sigma _{x}^{2}\smallskip & \rho \sigma _{x}\sigma _{p} \\ 
\rho \sigma _{x}\sigma _{p} & \sigma _{p}^{2}%
\end{bmatrix}%
\end{equation*}%
is sometimes said to be squeezed if either $\sigma _{x}^{2}$ or $\sigma
_{p}^{2}$ is smaller that $\frac{1}{2}\hbar $.\ However, as noticed and
discussed in Arvind et al. \cite{ADMS}, such a definition has no interesting
or useful invariance property; in particular it is no more preserved by
symplectic transformations than is the \textquotedblleft
popular\textquotedblright\ form $\sigma _{x}\sigma _{p}\geq \frac{1}{2}\hbar 
$ of the uncertainty principle. This motivates the following definition (%
\textit{ibid}.), valid for any number of degrees of freedom:

\begin{center}
\emph{A state is said to be squeezed if the smallest eigenvalue of the
covariance matrix }$\Sigma $\emph{\ is less than }$\frac{1}{2}\hbar $\emph{.}
\end{center}

The condition above is clearly $\limfunc{U}(n)$-invariant ($U^{T}\Sigma U$
has the same eigenvalues as $\Sigma $ for every $U\in \limfunc{U}(n)$), but
it is not $\limfunc{Sp}(n)$-invariant: we can thus use non-unitary elements
of the symplectic group to change the squeezed or non-squeezed status of a
quantum state. Let us discuss this from the point of view of admissible
ellipsoids. First, all quantum blobs which are not a ball correspond to a
squeezed state. Let in fact $\mathbb{Q}=S(B^{2n}(\sqrt{\hbar }))$; in view
of Lemma \ref{unit} for every $S\in \limfunc{Sp}(n)$ we can find $U\in 
\limfunc{U}(n)$ such that $SS^{T}=U^{T}\Delta U\ $where%
\begin{equation*}
\Delta =%
\begin{bmatrix}
\Lambda & 0 \\ 
0 & \Lambda ^{-1}%
\end{bmatrix}%
\text{ \ , \ \ }\Lambda =\limfunc{diag}[\lambda _{1},\lambda
_{2},...,\lambda _{n}]
\end{equation*}%
the $\lambda _{j}$, $1\leq j\leq n$ being the first $n$ eigenvalues
eigenvalues of $SS^{T}$; in particular $0<\lambda _{j}\leq 1$. The
corresponding covariance matrix is 
\begin{equation*}
\Sigma =\frac{\hbar }{2}SS^{T}=\frac{\hbar }{2}U^{T}\Delta U\ 
\end{equation*}%
whose eigenvalues are the numbers 
\begin{equation*}
\frac{\hbar }{2}\lambda _{1}\leq \cdot \cdot \cdot \leq \frac{\hbar }{2}%
\lambda _{n}\leq \frac{\hbar }{2}\leq \frac{\hbar }{2\lambda _{n}}\leq \cdot
\cdot \cdot \leq \frac{\hbar }{2\lambda _{1}}
\end{equation*}%
The only possibility for the state $\emph{not}$ to be squeezed is thus when
all the $\lambda _{j}$ are equal to one, and this corresponds to the trivial
case $\mathbb{Q}=B^{2n}(\sqrt{\hbar })$. More generally, Proposition \ref%
{bon} shows that:

\begin{center}
\emph{An ellipsoid }$\mathbb{B}:z^{T}Fz\leq \hbar $\emph{\ represents a
squeezed quantum state if and only if the eigenvalues of }$JF$\emph{\ are }$%
\leq 1$\emph{\ and the largest eigenvalue of }$F$\emph{\ is }$>1$\emph{.}
\end{center}

\section{Quantum Blobs and Wigner functions\label{four}}

As announced in the Introduction, there is a one-to-one correspondence
between quantum blobs and Gaussian pure states. Perhaps even more
interestingly, quantum mechanically admissible ellipsoids correspond to
Gaussian mixed states, and if such an admissible ellipsoid has symplectic
capacity exactly one-half of the quantum of action we can associate to it a
canonical \textquotedblleft companion Gaussian pure state\textquotedblright .

Let us begin by recalling some basic results about the Wigner transform (for
proofs and details see for instance Littlejohn \cite{Littlejohn};
interesting additional results can be found in Folland \cite{Folland} (the
latter however uses different normalizations); also see Grossmann \textit{et
al}. \cite{Grossmann}, and \cite{Berry1,Berry2,Ozorio} for applications to
integrable and non-integrable systems.

The Wigner transform $W\Psi $ of a function $\Psi \in L^{2}(\mathbb{R}%
_{x}^{n})$ is defined by%
\begin{equation}
W\Psi (z)=\left( \tfrac{1}{2\pi \hbar }\right) ^{n}\int \exp \left( -\tfrac{%
\mathrm{i}}{\hbar }py\right) \Psi (x+\tfrac{1}{2}y)\Psi ^{\ast }(x-\tfrac{1}{%
2}y)\mathrm{d}^{n}y  \label{wig}
\end{equation}%
and if $\Psi \in L^{1}(\mathbb{R}_{x}^{n})\cap L^{2}(\mathbb{R}_{x}^{n})$
then 
\begin{equation*}
\int W\Psi (z)\mathrm{d}^{n}p=|\Psi (x)|^{2}\text{ \ \ , \ \ }\int W\Psi (z)%
\mathrm{d}^{n}x=|\widehat{\Psi }(p)|^{2}
\end{equation*}%
where $\hat{\Psi}$ is the quantum Fourier transform: 
\begin{equation}
\hat{\Psi}(p)=\left( \tfrac{1}{2\pi \hbar }\right) ^{n/2}\int \exp \left( -%
\tfrac{\mathrm{i}}{\hbar }px\right) \Psi (x)\mathrm{d}^{n}x.  \label{quft}
\end{equation}%
The Wigner transform is a real function; however $W\Psi \geq 0$ if and only
if is $\Psi $ is a Gaussian (Hudson \cite{Hudson}); it is therefore not a
true phase-space probability but only a \textquotedblleft quasi probability
distribution\textquotedblright\ (see the discussion in Simon \textit{et al}. 
\cite{SSM}). We will use the following properties of $W\Psi $:

\begin{itemize}
\item Translations:%
\begin{equation}
W\Psi (z-z_{0})=W(\widehat{T}(z_{0})\Psi )(z)  \label{trans}
\end{equation}%
where $\widehat{T}(z_{0})$ is the Heisenberg operator defined by 
\begin{equation}
\widehat{T}(z_{0})\Psi (x)=\exp \left[ \tfrac{\mathrm{i}}{\hbar }(p_{0}x-%
\tfrac{1}{2}p_{0}x_{0})\right] \Psi (x-x_{0})\text{.}  \label{hey}
\end{equation}

\item Metaplectic covariance: for every $S\in \limfunc{Sp}(n)$%
\begin{equation}
W\Psi (S^{-1}z)=W(\hat{S}\Psi )(z)  \label{meco}
\end{equation}%
where $\hat{S}$ is any of the two operators $\pm \hat{S}\in Mp(n)$ ($Mp(n)$
is the metaplectic group) associated with $S$.
\end{itemize}

The Wigner transform is not invertible; however 
\begin{equation}
W\Psi =W\Psi ^{\prime }\Longleftrightarrow \Psi =c\Psi ^{\prime }\text{ \ ,
\ }|c|=1  \label{c}
\end{equation}%
so that the $W\Psi $ determines $\Psi $only up to an overall phase.

Before we study the fundamental correspondence between quantum blobs and
Gaussians, let us prove a technical result.

\subsection{A reduction result\label{factor}}

It turns out that we do not need all the elements of $\limfunc{Sp}(n)$ to
generate the set $\limfunc{Quant}_{0}(n)$ of quantum blobs centered at $%
z_{0}=0$. This has already been proved above, where we showed that all
quantum blobs can be obtained from the ball $B^{2n}(\sqrt{\hbar })$ by
translations, rescalings, and rotations. We are going to prove here another
\textquotedblleft reduction result\textquotedblright , taking into account
the fact that the ball $B^{2n}(\sqrt{\hbar })$ is insensitive to the action
of $\limfunc{U}(n)$. (This is of course already reflected by the
identification $\limfunc{Quant}_{0}(n)\equiv \limfunc{Sp}(n)/\limfunc{U}(n)$%
.)

Let $S$ be a symplectic matrix in block-form%
\begin{equation}
S=%
\begin{bmatrix}
A & B \\ 
C & D%
\end{bmatrix}%
\text{.}  \label{bloc}
\end{equation}

\begin{lemma}
\label{fac}Every symplectic matrix (\ref{bloc}) can be written, in a unique
way, as a product 
\begin{equation}
S=%
\begin{bmatrix}
A_{0} & 0 \\ 
C_{0} & A_{0}^{-1}%
\end{bmatrix}%
\begin{bmatrix}
X_{0} & -Y_{0} \\ 
Y_{0} & X_{0}%
\end{bmatrix}%
\text{.}  \label{iwa}
\end{equation}%
where $A_{0}$, $C_{0}$, $X_{0}$ and $Y_{0}$ are real $n\times n$ matrices
such that%
\begin{equation*}
A_{0}>0,\text{ }A_{0}C_{0}=C_{0}^{T}A,\text{ and }X_{0}+\mathrm{i}Y_{0}\in 
\limfunc{U}(n);
\end{equation*}%
when $S$ is written in block form (\ref{bloc}) these matrices are given by
the formulas%
\begin{align}
A_{0}& =(AA^{T}+BB^{T})^{1/2}  \label{1} \\
C_{0}& =(CA^{T}+DB^{T})(AA^{T}+BB^{T})^{-1/2}  \label{2} \\
X_{0}+\mathrm{i}Y_{0}& =(AA^{T}+BB^{T})^{-1/2}(A-\mathrm{i}B).  \label{3}
\end{align}
\end{lemma}

\begin{proof}
The unitary group $\limfunc{U}(n)$ acts transitively on the Lagrangian
Grassmannian ((i.e. the set of all Lagrangian planes). This implies that
there exists $U\in \limfunc{U}(n)$ such that $S\ell _{p}=U\ell _{p}$ (recall
that $\ell _{p}=0\times \mathbb{R}_{p}^{n}$) and hence $S=RU$ for some $R\in 
\limfunc{Sp}(n)$ such that $R\ell _{p}=\ell _{p}$. This implies that $R$
must be of the form 
\begin{equation*}
R=%
\begin{bmatrix}
A_{0} & 0 \\ 
C_{0} & A_{0}^{-1}%
\end{bmatrix}%
\end{equation*}%
hence (\ref{iwa}). Let us show that $AA^{T}+BB^{T}$ is invertible; formulae (%
\ref{1}), (\ref{2}), (\ref{3}) then easily follow using the relations
imposed on $A,B,C,$ and $D$ by the fact that $S$ is symplectic. Expanding
the product (\ref{iwa}) we get%
\begin{equation*}
A=A_{0}X_{0}\text{ \ , \ }B=-A_{0}Y_{0}
\end{equation*}%
and hence%
\begin{equation*}
AA^{T}+BB^{T}=A_{0}(X_{0}X_{0}^{T}+Y_{0}Y_{0}^{T})A_{0}^{T}=A_{0}A_{0}^{T}
\end{equation*}%
that is, since $A_{0}$ is invertible, 
\begin{equation*}
\det (AA^{T}+BB^{T})\neq 0
\end{equation*}%
as claimed.
\end{proof}

It follows that:

\begin{corollary}
Every quantum blob $\mathbb{Q}\in \limfunc{Quant}_{0}(n)$ is the image of
the ball $B^{2n}(\sqrt{\hbar })$ by a symplectic transformation of the type%
\begin{equation*}
S=%
\begin{bmatrix}
A_{0} & 0 \\ 
C_{0} & A_{0}^{-1}%
\end{bmatrix}%
\end{equation*}%
with $A_{0}>0$ and $A_{0}C_{0}=C_{0}^{T}A$.
\end{corollary}

\begin{proof}
It is an obvious consequence of Lemma \ref{fac} since $U(B^{2n}(\sqrt{\hbar }%
))=B^{2n}(\sqrt{\hbar })$ for every $U\in \limfunc{U}(n)$.
\end{proof}

\subsection{The Fundamental Correspondence}

Let us take a close scrutiny on the relationship between quantum blobs (and,
more generally, admissible ellipsoids) and Wigner transforms.

We will denote by $\limfunc{Gauss}_{0}(n)$ the set of all centered and
normalized Gaussian functions given by

\begin{equation}
\Psi (x)=\left( \tfrac{\det X}{(\pi \hbar )^{n}}\right) ^{1/4}\exp \left[ -%
\tfrac{1}{2\hbar }x^{T}(X+\mathrm{i}Y)x\right]  \label{gauss}
\end{equation}%
where $X$ and $Y$ are real symmetric $n\times n$ matrices, $X>0$. The Wigner
transform $W\Psi $ of such a function is the phase space Gaussian%
\begin{equation}
W\Psi (z)=\left( \tfrac{1}{\pi \hbar }\right) ^{n}\exp \left( -\tfrac{1}{%
\hbar }z^{T}Gz\right)  \label{wpsi}
\end{equation}%
where $G$ is the matrix%
\begin{equation}
G=%
\begin{bmatrix}
X+YX^{-1}Y & YX^{-1} \\ 
X^{-1}Y & X^{-1}%
\end{bmatrix}
\label{g}
\end{equation}%
(Littlejohn, \cite{Littlejohn}). A fundamental remark is that $G$ is both
symmetric positive definite and \emph{symplectic}. The latter property is
actually straightforward, checking that $G^{T}JG=J,$ and the positive
definiteness is seen by noting that if $z\neq 0$ then, setting $y=Yx+p$,%
\begin{equation*}
z^{T}Gz=x^{T}Xx+y^{T}X^{-1}y>0\text{.}
\end{equation*}%
The associated covariance matrix is defined, as in Subsection \ref{reluc}, by%
\begin{equation*}
\Sigma =\frac{\hbar }{2}G^{-1}.
\end{equation*}

We next notice that an immediate calculation shows that $G$ can be factored
as follows:%
\begin{equation}
G=(SS^{T})^{-1}\text{ \ \ with \ \ }S=%
\begin{bmatrix}
X^{-1/2} & 0 \\ 
-YX^{-1/2} & X^{1/2}%
\end{bmatrix}
\label{bi}
\end{equation}%
hence the set $\mathbb{Q}:z^{T}(SS^{T})^{-1}z\leq \hbar $ is the quantum
blob $\mathbb{Q}=S(B^{2n}(\sqrt{\hbar }))$. Thus, to a Gaussian $\Psi $ we
have associated in a canonical way a quantum blob $\mathbb{Q}=\mathbb{Q}%
(\Psi )$. The remarkable fact is that this correspondence is a bijection:

\begin{theorem}
\label{theo}The mapping $\mathbb{Q}(\cdot ):\limfunc{Gauss}%
\nolimits_{0}(n)\longrightarrow \limfunc{Quant}\nolimits_{0}(n)$ which to
every centered Gaussian $\Psi $ associates the quantum blob $\mathbb{Q}(\Psi
)=S(B^{2n}(\sqrt{\hbar }))$ where $S$ is defined by (\ref{bi}) is a
bijection.
\end{theorem}

\begin{proof}
Assume that $\mathbb{Q}\in \limfunc{Quant}\nolimits_{0}(n)$ is given by $%
\mathbb{Q}=S(B^{2n}(\sqrt{\hbar })),$ that is $z^{T}(SS^{T})^{-1}z\leq \hbar 
$; in view of Lemma \ref{fac} in Subsection \ref{factor} there exist $%
A_{0}>0 $ and $C_{0}$ such that $A_{0}C_{0}=C_{0}^{T}A_{0}$ and%
\begin{eqnarray*}
SS^{T} &=&%
\begin{bmatrix}
A_{0} & 0 \\ 
C_{0} & A_{0}^{-1}%
\end{bmatrix}%
\begin{bmatrix}
A_{0} & C_{0}^{T} \\ 
0 & A_{0}^{-1}%
\end{bmatrix}
\\
&=&%
\begin{bmatrix}
A_{0}^{2} & A_{0}C_{0}^{T} \\ 
C_{0}A_{0} & C_{0}C_{0}^{T}+A_{0}^{-2}%
\end{bmatrix}%
\end{eqnarray*}%
and hence%
\begin{equation*}
(SS^{T})^{-1}=%
\begin{bmatrix}
C_{0}C_{0}^{T}+A_{0}^{-2} & -C_{0}A_{0} \\ 
-A_{0}C_{0}^{T} & A_{0}^{2}%
\end{bmatrix}%
\text{.}
\end{equation*}%
The equation%
\begin{equation*}
\begin{bmatrix}
X+YX^{-1}Y & YX^{-1} \\ 
X^{-1}Y & X^{-1}%
\end{bmatrix}%
=%
\begin{bmatrix}
C_{0}C_{0}^{T}+A_{0}^{-2} & -C_{0}A_{0} \\ 
-A_{0}C_{0}^{T} & A_{0}^{2}%
\end{bmatrix}%
\end{equation*}%
has the unique solutions%
\begin{equation*}
X=A_{0}^{-2}\text{ \ , \ }Y=-A_{0}^{-1}C_{0}^{T}\text{.}
\end{equation*}
\end{proof}

\begin{remark}
This result shows that there is a bijective correspondence between the set $%
\Sigma (n)$ all complex symmetric $n\times n$ matrices with definite
positive real part (\textquotedblleft Siegel half-plane\textquotedblright )
and $\limfunc{Sp}(n)/\limfunc{U}(n).$ This has also been noticed in \cite%
{Folland} in a different context.
\end{remark}

The extension of Theorem \ref{theo} to the case of Gaussians with an
arbitrary center is straightforward and left to the reader; on establishes
the existence of a bijection%
\begin{equation*}
\mathbb{Q}(\cdot ):\limfunc{Gauss}(n)\longrightarrow \limfunc{Quant}(n)
\end{equation*}%
between the set of all Gaussians obtained from (\ref{gauss}) by translation
and the set of all quantum blobs.

Another remarkable fact is that we can associate in a canonical manner a
\textquotedblleft companion coherent state\textquotedblright\ to every
quantum mechanically admissible ellipsoid with minimum capacity $\frac{1}{2}%
h $. In this sense, those ellipsoids appear as \textquotedblleft quasi
quantum-blobs\textquotedblright .

\begin{corollary}
\label{coro}Let $\mathbb{B}$ be an admissible ellipsoid with $\underline{c}(%
\mathbb{B})=\frac{1}{2}h$. Then there exists a unique centered Gaussian (\ref%
{gauss}), denoted by $\Psi _{\mathbb{B}}$, such that%
\begin{equation}
W\Psi _{\mathbb{B}}(z)=\left( \tfrac{1}{\pi \hbar }\right) ^{n}\exp \left( -%
\tfrac{1}{\hbar }z^{T}Gz\right) \text{ \ , \ }G=S^{T}S  \label{companion}
\end{equation}%
where $S\in \limfunc{Sp}(n)$ is any symplectic matrix such that $S(B^{2n}(%
\sqrt{\hbar }))\subset \mathbb{B}$.
\end{corollary}

\begin{proof}
We must check that the right-hand side of (\ref{companion}) is independent
of the choice of the symplectic matrix $S$ putting\ $M$ in the Williamson
diagonal form. Let $S$ and $S^{\prime }$ be two such choices; in view of
Lemma \ref{us} (Section \ref{one}) there exists $U\in \limfunc{U}(n)$ such
that $S=US^{\prime }$ and hence 
\begin{equation*}
z^{T}S^{T}Sz=z^{T}S^{\prime T}U^{T}US^{\prime }zz^{T}=z^{T}S^{\prime
T}S^{\prime }z
\end{equation*}%
whence the result.
\end{proof}

The generalization of the two results above to Gaussians and admissible sets
centered at arbitrary points is straightforward using the property (\ref%
{trans}) of Wigner transforms. We leave it to the reader to restate Theorem %
\ref{theo} and its Corollary \ref{coro} in this more general setting.

\subsection{Averaging on quantum blobs}

Assume that $\widehat{\rho }$ is some arbitrary (mixed) state: $Tr(\widehat{%
\rho }^{2})\leq 1$. By the usual Weyl calculus \cite{Folland,Littlejohn} one
can associate to $\widehat{\rho }$ it Wigner transform $W\widehat{\rho }$.
The state $\widehat{\rho }$ is said to be \emph{Gaussian} if there exists $%
F>0$ such that%
\begin{equation}
W\widehat{\rho }(z)=\left( \tfrac{1}{\pi \hbar }\right) ^{n}(\det
F)^{1/2}\exp \left( -\tfrac{1}{\hbar }z^{T}Fz\right) ;  \label{wro}
\end{equation}%
the eigenvalues of the covariance matrix $\Sigma =\frac{\hbar }{2}F^{-1}$
are all $\geq \frac{1}{2}\hbar $. This is precisely condition (\ref{unce}%
)(C) in Proposition \ref{bon}, thus:

\begin{center}
\emph{A phase space function (\ref{wro}) is the Wigner transform of a
(mixed) quantum state if and only if the ellipsoid }$\mathbb{B}:z^{T}Fz\leq
\hbar $\emph{\ is admissible. }
\end{center}

Consider now a quite general Gaussian of the type%
\begin{equation}
W(z)=\left( \tfrac{1}{\pi \hbar }\right) ^{n}(\det H)^{1/2}\exp \left( -%
\tfrac{1}{\hbar }z^{T}Hz\right)  \label{wz}
\end{equation}%
with $H>0$. We are going to see that if we \textquotedblleft
average\textquotedblright\ that function $W$ over a quantum blob (in a sense
that will be made precise below), then we will always obtain the Wigner
transform of a (mixed) Gaussian state.

\begin{theorem}
\label{hus}Let $\mathbb{Q}=S(B^{2n}(\sqrt{\hbar }))$ be a quantum blob and $%
\Psi _{\mathbb{Q}}$ the associated Gaussian (\ref{gauss}). The convolution
product $W_{\mathbb{Q}}=W\ast W\Psi _{\mathbb{Q}}$ where $W$ is any Gaussian
(\ref{wz}) is the Wigner transform of some mixed state $\widehat{\rho }$. In
fact,%
\begin{equation}
W\widehat{\rho }(z)=\left( \tfrac{1}{\pi \hbar }\right) ^{n}(\det
F)^{1/2}\exp \left( -\tfrac{1}{\hbar }z^{T}Fz\right)  \label{fi}
\end{equation}%
where the symplectic spectrum of $F$ is the image of that of $F$ by the
mapping $\lambda \longmapsto \lambda /(1+\lambda )$.
\end{theorem}

\begin{proof}
We have, by definition,%
\begin{equation*}
W\Psi _{\mathbb{Q}}(z)=\left( \tfrac{1}{\pi \hbar }\right) ^{n}\exp \left( -%
\tfrac{1}{\hbar }z^{T}Gz\right) \text{ \ , }G>0\text{ \ , \ }G\in \limfunc{Sp%
}(n)
\end{equation*}%
where $G=(SS^{T})^{-1}$ hence%
\begin{equation*}
W_{\mathbb{Q}}(z)=\left( \tfrac{1}{\pi \hbar }\right) ^{2n}(\det
F)^{1/2}\int e^{-\tfrac{1}{\hbar }(z-z^{\prime })^{T}H(z-z^{\prime })}e^{-%
\tfrac{1}{\hbar }(z^{\prime })^{T}Gz^{\prime }}\mathrm{d}^{2n}z^{\prime }%
\text{.}
\end{equation*}%
Setting $z^{\prime \prime }=S^{-1}z^{\prime }$ we have%
\begin{equation*}
W_{\mathbb{Q}}(Sz)=\left( \tfrac{1}{\pi \hbar }\right) ^{2n}(\det
H)^{1/2}\int e^{-\tfrac{1}{\hbar }(z-z^{\prime \prime
})^{T}S^{T}HS(z-z^{\prime \prime })}e^{-\tfrac{1}{\hbar }|z^{\prime \prime
}|^{2}}\mathrm{d}^{2n}z^{\prime \prime }\text{.}
\end{equation*}%
Replacing if necessary $S$ by another symplectic matrix we may assume, in
view of Williamson's theorem, that%
\begin{equation*}
S^{T}HS=D=%
\begin{bmatrix}
\Lambda _{\omega } & 0 \\ 
0 & \Lambda _{\omega }%
\end{bmatrix}%
\end{equation*}%
and hence%
\begin{equation*}
W_{\mathbb{Q}}(Sz)=\left( \tfrac{1}{\pi \hbar }\right) ^{2n}(\det
D)^{1/2}\int e^{-\tfrac{1}{\hbar }(z-z^{\prime \prime })^{T}D(z-z^{\prime
\prime })}e^{-\tfrac{1}{\hbar }|z^{\prime \prime }|^{2}}\mathrm{d}%
^{2n}z^{\prime \prime }\text{.}
\end{equation*}%
Using the elementary formula%
\begin{equation*}
\int_{-\infty }^{\infty }e^{-a(u-t)^{2}}e^{-bt^{2}}\mathrm{d}t=\sqrt{\frac{%
\pi }{a+b}}\exp \left( -\frac{ab}{a+b}u^{2}\right)
\end{equation*}%
valid for all $a,b>0$ together with the fact that the matrix $D$ is diagonal
we find%
\begin{equation*}
W_{\mathbb{Q}}(Sz)=\left( \tfrac{1}{\pi \hbar }\right) ^{n}(\det
D(I+D)^{-1})^{1/2}\exp \left[ -\tfrac{1}{\hbar }z^{T}D(I+D)^{-1}z\right]
\end{equation*}%
that is%
\begin{equation*}
W_{\mathbb{Q}}(z)=\exp \left[ -\tfrac{1}{\hbar }%
z^{T}(S^{-1})^{T}D(I+D)^{-1}S^{-1}z\right] .
\end{equation*}%
Setting 
\begin{equation*}
F=(S^{-1})^{T}D(I+D)^{-1}S^{-1}
\end{equation*}%
we obtain formula (\ref{fi}). There remains to show that the moduli of the
eigenvalues of $JF$ are not superior to one. Since $S\in \limfunc{Sp}(n)$ we
have $J(S^{-1})^{T}=SJ$ and hence 
\begin{equation*}
JF=SJD(I+D)^{-1}S^{-1}
\end{equation*}%
has the same eigenvalues as $JD(I+D)^{-1}$. Now,%
\begin{equation*}
JD(I+D)^{-1}=%
\begin{bmatrix}
0 & \Lambda _{\omega }(I+\Lambda _{\omega })^{-1} \\ 
-\Lambda _{\omega }(I+\Lambda _{\omega })^{-1} & 0%
\end{bmatrix}%
\end{equation*}%
so that the eigenvalues $\mu _{\omega ,j}$ of $JF$ are the numbers%
\begin{equation}
\mu _{\omega ,j}=\pm \mathrm{i}\frac{\lambda _{\omega ,j}}{1+\lambda
_{\omega ,j}}  \label{mugi}
\end{equation}%
which are such that $|\mu _{\omega ,j}|\leq 1$ which was to be proven.
\end{proof}

This result is reminiscent of (and related to) the well-known property of
the Husimi transform (Husimi \cite{Husimi}). It is optimal in the following
sense: let $\alpha $ and $\beta $ be two positive numbers and set 
\begin{equation*}
\Phi _{\alpha ,\beta }(z)=\exp \left[ -\tfrac{1}{\hbar }\left( \tfrac{|x|^{2}%
}{\alpha ^{2}}+\tfrac{|p|^{2}}{\beta ^{2}}\right) \right] \text{.}
\end{equation*}%
As de Bruijn \cite{Bruijn} has shown, the convolution product $W\Psi \ast
\Phi _{\alpha ,\beta }$ is always positive as long as $\alpha \beta \geq 1$,
but it can take negative values when $\alpha \beta <1$. Now, the ellipsoid%
\begin{equation*}
\mathbb{B}:\frac{|x|^{2}}{\alpha ^{2}}+\frac{|p|^{2}}{\beta ^{2}}\leq \hbar
\end{equation*}%
is the image of the ball $B^{2n}(\alpha \beta \sqrt{\hbar })$ by the
symplectic transformation%
\begin{equation*}
(x,p)\longmapsto (x\sqrt{\alpha /\beta },p\sqrt{\beta /\alpha })\text{;}
\end{equation*}%
$\mathbb{B}$ is thus a quantum mechanically admissible ellipsoid if and only
if $\alpha \beta \geq 1$.

\section{Concluding Remarks and Perspectives}

We have presented a general framework for the study of theoretical phase
space properties associated to quantum mechanical systems. The fact that our
treatment is essentially linear is not \textit{per se} a real limitation
because the uncertainty principle is only invariant under linear canonical
transformations. On the other hand, it is also well-known that linear
symplectic transformations correspond to those operations that preserve the
Gaussian character of states and that can be implemented by means of optical
elements such as beam splitters, phase shifts, and squeezers together with
homodyne measurements (see for instance the discussion in Giedke \textit{et
al}. \cite{GECP}). Still, it would be interesting, at least from a
mathematical viewpoint, to have a closer look on non-linear blobs, images by
arbitrary canonical transformations of the ball $B^{2n}(\sqrt{\hbar })$. One
step in that direction is to note that while Theorem \ref{un} probably does
not remain true for non-linear blobs, its Corollary \ref{coco} does:
Gromov's non-squeezing theorem namely says that the area of the orthogonal
projection of a ball $f(B^{2n}(r))$ $(f$ a canonical transformation) on any
symplectic plane is always at least $\pi r^{2}$. It would be interesting to
see whether one could deduce from this property non-linear analogues of
Proposition \ref{bon}, Theorem \ref{theo}, and its Corollary \ref{coro}.

It would certainly also be interesting, and perhaps useful, to recast our
constructions in the context of Howe's \textquotedblleft oscillator
semigroup\textquotedblright\ calculus \cite{Howe} (reviewed in the last
Chapter in Folland \cite{Folland}); Gaussians playing a fundamental role in
both theories seems to indicate that there might be some intimate connection
between quantum blobs and the Gaussian kernel operators introduced by Howe.

But perhaps the most fashionable --and interesting!-- application of our
methods would still be the study of entanglement of multi-partite systems,
so essential in the understanding of various EPR-type phenomena like
teleportation. What makes us believe that the notion of quantum blob could
play a pivotal role in this question is that, as we already pointed out in
the Introduction, quantum blobs have an arbitrarily large phase space
extension, which makes them good tools to use in questions involving
non-locality. We will return to these so fascinating and important questions
in a near future.

\begin{acknowledgement}
I wish to thank Michael Hall (private communication) for having suggested to
me the relation between symplectic capacities, Gaussians, and the Wigner
transform.
\end{acknowledgement}

\end{document}